# Berry phase induced localization to anti-localization transition in two-dimensional Dirac fermion systems

*Ting Zhang*[1,2], *Jie Pan*[1], *Ping Sheng*[1,2]

[1]Department of Physics, Hong Kong University of Science and Technology
Clear Water Bay, Kowloon, Hong Kong, China
and
[2]Institute for Advanced Study, Hong Kong University of Science and Technology
Clear Water Bay, Kowloon, Hong Kong, China

**Abstract**

We study theoretically the electrical transport of two-dimensional (2D) massive Dirac fermions, which are described by the Hamiltonian $H_0 = \upsilon \vec{\sigma} \cdot \vec{k} + m\sigma_z$, and with a gap at the charge neutrality point. Through analytical diagrammatical calculations of electrical conductivity in the presence of long range Coulomb scattering centers, we show that attendant with the variation of the Berry phase from 0 to $\pi$ as the Fermi energy moves away from the Dirac point/band boundary, a continuous Anderson-localization (AL) to weak-localization (WL), and further to weak anti-localization (WAL) transition occurs, implying a change in the sign of the magnetoresistance. Such transition indicates the presence of metal-insulator transition (MIT) in this 2D system, in contrast to the classical scaling theory. The WL to WAL transition occurs at a certain critical Berry phase despite the concentration of Coulomb impurities, while the MIT critical point, which is the distinguishing doping level separating the AL and WL phases, depends on the competition of conventional conductivity and the negative maximally crossed diagram (MCD) corrections near the bottom of conduction band.

Localization behavior [1,2], which denotes the conductivity of a disordered system exponentially decays as system size increases, is widely observed in disordered solid state materials. It is a very important transport phenomenon in realistic systems, since disorders and impurities are always present, and therefore influences the performance of electric devices. Localization properties are strongly dependent on the sample's dimensionality, i.e. according to the famous scaling theory [3-5] of localization, the function describing the variation of conductivity with respect to sample size is $\beta = d\ln(g)/d\ln L \sim d - 2$ in the high mobility limit, whereas $g$ is the conductivity of the sample, $L$ is the size scale, and $d$ means the dimension. In 3-dimension ($d=3$) it is positive in this limit, means the existence of extended state and metallic phase, and if the strength of disorders is very high, i.e. in the other localized limit, $\beta$ could also be negative, identifying a localization phase. In 3D, $\beta=0$ defines a mobility edge, which distinguishes the extended and localization phases. However in 2D and 1D, $\beta$ should be always negative and no extended states should present. The scaling theory describes transportation in conventional disordered electronic systems quite well, however in recent years un-conventional 2D materials like graphene [6-8] show different behaviors. Physically, propagating waves scattered on the randomly distributed impurities and interfere with each other. In the presence of time-reversal symmetry, for any closed propagation path, the forward and backward propagating waves will constructively interfere and backscattering is enhanced [5,9-12]. This so-called coherent backscattering process [5,9-12] give rise to localization in conventional electronic systems. However in graphene based materials satisfying 2D Weyl equation, the associated binary wave-functions contain a Berry phase π. This Berry phase introduces weak anti-localization (WAL) behaviors [13-15] in graphene, since in this case the backscattering is suppressed by this additional phase factor [16-20]. In earlier works [16-20] researchers identified this WAL phase and investigated the WAL to WL transition with respect to different disorder strengths [21] or different disorder types [16-20,22,23]. However, since in a certain sample the disorder type or strengths are both fixed, such transition is hard to observe in experiment. In this manuscript, we show another possibility, that the AL/WL to WAL transition can be realized in the same 2D sample, with respect to different doping level. We consider the 2D massive Dirac Hamiltonian [25-28]:

$$H_0 = \upsilon \vec{\sigma} \cdot \vec{k} + m\sigma_z \qquad (1)$$

where $\upsilon$ denotes the group velocity of electrons, $\vec{\sigma} = (\sigma_x, \sigma_y)$ is the Pauli matrices in sub-lattice space, $\vec{k}$ is the 2D momentum vector measured from the K or K' point. In eq.1 the most distinguishing character from the Weyl equation is the $m\sigma_z$ term, which indicates an asymmetric interaction potential between the different sub-lattice sites. This mass term in Hamiltonian introduces a gap in the energy spectrum, around the charge neutrality point. It could be realized in certain graphene/substrate composites, which may induce A/B asymmetry [28]. As suggested by R. Skomski et. al. [28], such sub-lattice asymmetry could occur in epitaxially grown graphene /hexagonal-boron nitride (*h*-BN) bilayer system where the graphene layer is in registry with the *h*-BN layer (mass gap ~53 *meV*) [29], in graphene on SiC (with a gap ~0.26 *eV*) [30,31], and in graphene on MgO (gap ~0.5−1 *eV*) [32-34] systems. Although in the following works graphene/ h-BN systems are always with Morrie super-lattices [35-37], A/B asymmetry still occur and the Hamiltonian in eq.1 are still hopeful to realize in some other systems.

In this work, based on Hamiltonian in eq. 1, we carry out analytical diagrammatic calculations on the Drude-conductivity and its corrections due to disorder scatterings by the long-range Coulomb scatterers [16-20,22-23,38-43]. We find a continuous AL/WL/WAL transition as the doping level moves away from the charge neutrality point (CNP), attendant with the increase of the Berry phase [16,44,45] from zero towards $\pi$. For the parameters of graphene/h-BN system substituted, the critical doping point for WL/WAL transition is identified to be ~46 meV from the bottom of the conduction band, in the limit of low impurity concentration. Furthermore, an AL/WL transition is identified in this 2D system as a function of impurity concentration, which is in contrast to the prediction of classical scaling theory. A phase diagram for AL/WL/WAL regions with different mass terms is calculated.

The energy dispersion for the Hamiltonian, eq. (1), is easily shown to be $\varepsilon = \pm\sqrt{v^2 k^2 + m^2}$, where a gap of $2m$ is obvious. For the eigen-functions, we have (in the conduction band (CB)):

$$\phi_{\vec{k},A} = \sqrt{\frac{\varepsilon + m}{2\varepsilon}} e^{i\vec{k}\cdot\vec{r}}, \tag{2a}$$

$$\phi_{\vec{k},B} = \sqrt{\frac{\varepsilon + m}{2\varepsilon}} \frac{v(k_x + ik_y)}{(\varepsilon + m)} e^{i\vec{k}\cdot\vec{r}}. \tag{2b}$$

They are noted to be quite different from each other. For the valance band the roles of A and B sites are reversed. The difference diminishes away from the bottom of the band. Due to the Fermi energy dependence of the wavefunction, the Berry phase also acquires energy dependence. The Berry curvature of the system is given by [44,45] $\Omega(k) = |\nabla_k \times \langle \psi | i\nabla_k | \psi \rangle| = \frac{m}{2\varepsilon^3}$, where $\psi = (\phi_{\vec{k},A}, \phi_{\vec{k},B})$ is the wavefunction. The Berry phase is obtained by integrating the Berry curvature: $\Lambda(k) = \oiint_S \Omega(k) d\vec{k} = \pi(\varepsilon - m)/\varepsilon$. Here $S$ is the area enclosed by the circle $|k| = k$. For $\varepsilon = m$, the bottom of the conduction band, $\Lambda(k) = 0$ and there is a smooth transition to $\Lambda(k) = \pi$ as $\varepsilon$ increases. This fact, which is encoded in the wavefunction of the system, will greatly influence the transport property of the A/B asymmetric system as seen below.

The single-particle Green's function can be expressed as:

$$G^{(0)R/A}(\vec{k},\varepsilon) = \frac{1}{\varepsilon - H_0} = \frac{\varepsilon + m\sigma_z + v\vec{\sigma}\cdot\vec{k}}{(\varepsilon \pm i\delta)^2 - (v^2 k^2 + m^2)} \sim \frac{1}{2\varepsilon} \frac{\varepsilon + m\sigma_z + v\vec{\sigma}\cdot\vec{k}}{\varepsilon - \sqrt{v^2 k^2 + m^2} \pm i\delta}. \tag{3}$$

Here we restrict ourselves to within a single valley, since only Coulomb scatterers are considered. We also limit ourselves to the CB, so $\varepsilon > 0$ as measured from the Dirac point. Randomly distributed long range disorders [38-43] are taken into account. Such long range disorders interact with electrons via long range Coulomb interaction, and give rise to scattering events within a single valley [38,39]. This behavior is quite different from the case of short range disorders, which would induce inter-valley scatterings in graphene and result in weak localization [17,23,46-63]. The correction to the Green function due to these long range disorder scatterings can be evaluated by the self-consistent Born approximation (SCBA) [64,65], in which the self-energy is given by:

$$\Sigma_{ij}(\vec{k},\varepsilon) = \int \frac{d\vec{k}'}{(2\pi)^2} \sum_{l,m=A,B} n_i u_{i,l}(\vec{k}-\vec{k}') G_{lm}(\vec{k}',\varepsilon) u_{m,j}(\vec{k}'-\vec{k}) \cdot \quad (4)$$

Here $\Sigma_{ij}(\vec{k},\varepsilon)$ is the matrix element of self-energy, with $i$, $j$, $l$ and $m$ run over the A/B sublattice atomic sites, $n_i$ denotes the density of impurities, and $u_{kk'AA} = u_{kk'BB} = 2\pi e^2/\kappa p\varepsilon(p)$, $u_{kk'AB} = u_{kk'BA} = 0$ denote the Coulomb potential in the momentum space, where $e$ is the electronic charge and $\kappa$ the dielectric constant of the environment, taken to be 2 in this work. Here $p = |\vec{k}'-\vec{k}|$ is the scattering momentum, with a magnitude given by $p = 2k\sin\varphi$, where $\varphi = (\theta'-\theta)/2$, and $\theta, \theta'$ denote the directional angles of the momenta $\vec{k}$ and $\vec{k}'$, respectively. Due to dynamic screening by the electron gas, the electronic dielectric function $\varepsilon(p)$ can be evaluated in the Thomas Fermi (TF) approximation [43,64], as a result, $p\varepsilon(p)$ may be expressed as $p + ip_{TF}$, where $p_{TF} = 2\pi e^2 N_0/\kappa$ is the TF momentum, and $N_0$ is the density of state [66]. The self-energy Dyson equation can be expressed graphically as shown in Fig. 1A. A detailed derivation is given in the Appendix A. The result of the calculation shows that the corrected Green function can be expressed as

$$G^{R/A}(\vec{k},\varepsilon) = \frac{1}{2\varepsilon} \frac{\varepsilon + m\sigma_z + \upsilon\vec{\sigma}\cdot\vec{k}}{\varepsilon - \sqrt{\upsilon^2 k^2 + m^2} \pm i/2\tau_k}, \tag{5a}$$

$$\frac{1}{\tau_k} = \left(\frac{2\pi e^2}{\kappa}\right)^2 \frac{n_i}{2\upsilon^2} \left(\frac{\varepsilon^2 + m^2}{\varepsilon} A_0 + \frac{\varepsilon^2 - m^2}{2\varepsilon}(A_2 + A_{-2})\right), \tag{5b}$$

where the angular integrals $A_{2n}$ ($n=0,1,2$) are given by

$$A_{2n} = \int \frac{2d\varphi}{2\pi} \frac{e^{i2n\varphi}}{\left(4k^2 \sin^2\varphi + q_{TF}^2\right)}, \tag{5c}$$

which can be evaluated numerically. The scattering is noted to be strongly anisotropic, which is in contrast to the earlier theoretical works [23] in which the scattering is isotropic for the short-range impurities.

Disorder scatterings also give rise to corrections to the current vertex in the calculation of the Drude conductivity, as shown in Fig. 1B. Such corrections can be expanded in series of multiple scattering events, as expressed in ladder diagrams. A self-consistent Bethe-Salpeter equation can be applied to describe this vertex correction, with a diagrammatical representation given in Fig. 1C:

$$J_x(\vec{k},\vec{q},\omega) = j_x + \int \frac{d\vec{k}'}{(2\pi)^2} n_i U_{k,k'} G^R(\vec{k}',\varepsilon) J_x(\vec{k}',\vec{q},\omega) G^A(\vec{k}'-\vec{q},\varepsilon-\omega) U_{k'-q,k-q}, \tag{6}$$

where $J_x$ denotes the corrected current vertex, $j_x = |\nabla_{\vec{k}} H|_x = \upsilon\sigma_x$ is the corresponding bare current vertex, and $U_{k,k'} = \begin{pmatrix} u_{kk'}^{(A)} & 0 \\ 0 & u_{kk'}^{(B)} \end{pmatrix}$ is the matrix of the disorder potential. Here $\vec{q}$ and $\omega$ are, respectively, the momentum and energy (frequency) difference between the retarded and advanced Green functions (denoted by superscripts $R$ and $A$, respectively). By decomposing the renormalized current vertex into a series of angular basis functions $e^{in\theta}$ with $n$ varies from -2 to 2, i.e., $J_x = J_x^{(0)} + J_x^{(1)}e^{i\theta} + J_x^{(-1)}e^{-i\theta} + J_x^{(2)}e^{i2\theta} + J_x^{(-2)}e^{-i2\theta}$, Eq. (6) can be written as a set of coupled self-consistent equations:

$$J_x^{(0)} + J_x^{(1)}e^{i\theta} + J_x^{(-1)}e^{-i\theta} + J_x^{(2)}e^{i2\theta} + J_x^{(-2)}e^{-i2\theta}$$
$$= j_x + \int \frac{d\vec{k}'}{(2\pi)^2} n_i U_{k,k'} G^R(\vec{k}',\varepsilon) \left( J_x^{(0)} + J_x^{(1)}e^{i\theta'} + J_x^{(-1)}e^{-i\theta'} + J_x^{(2)}e^{i2\theta'} + J_x^{(-2)}e^{-i2\theta'} \right) G^A(\vec{k}'-\vec{q},\varepsilon-\omega) U_{k'-q,k-q}$$
(7a)

Here the Green functions are noted to be a function of both $\theta$ and $\theta'$. Hence after the integration over $k'$, each $J_x^{(n)}$ ($n$=-2,-1,0,1,2) on the left hand side will be expressed in terms of the others, forming a coupled set of equations. By evaluating the integrals

$$\xi_{i,j,l,m}^n = \int \frac{d\vec{k}'}{(2\pi)^2} n_i U_{k,k',i,i} G^R_{i,l}(\vec{k}',\varepsilon) e^{in\theta'} G^A_{m,j}(\vec{k}'-\vec{q},\varepsilon-\omega) U_{k'-q,k-q,j,j} \quad , \tag{7b}$$

where $i$, $j$, $l$, $m$ run over the A/B sub-lattice sites, the equation set (7a) can be solved analytically with the help of Matlab. The detailed calculation of the integrals and the solution of the Bethe-Salpeter equation, Eq. (7a), are summarized in Appendix B. We find that in the high energy region, our result on the vertex correction is the same as that obtained by the earlier theoretical work on pristine graphene [23]. Near the band bottom, however, the correction goes to 0, which is the consequence of the gapped energy spectrum. In this case, we find the most important difference with conventional two dimensional electron gases (2DEG) is the lack of diffusion pole in this ladder diagram correction.

Now we can estimate the Drude conductivity up to the ladder diagram level. The diagram representation of the Drude conductivity is shown in Fig. 1A–Fig. 1C. The equation form is given by the sum of the zeroth and first order corrections as follows:

$$\sigma_{xx} = \frac{e^2}{\omega} \int_{\mu-\omega}^{\mu} \frac{d\varepsilon}{2\pi} Tr \int \frac{d\vec{k}}{(2\pi)^2} j_{k,x} G^r_k j_{k,x} G^a_k + \frac{e^2}{\omega} \int_{\mu-\omega}^{\mu} \frac{d\varepsilon}{2\pi} Tr \int \frac{d\vec{k}}{(2\pi)^2} \int \frac{d\vec{k}'}{(2\pi)^2} j_{k,x} G^r_k \left( n_i U_{kk'} G^r_{k'} J_{k',x} G^a_{k'} U_{k'k} \right) G^a_k, \quad (8)$$

where the upper limit in the energy integrals are noted to be the chemical potential $\mu$ (the Fermi level). Substituting the corresponding terms into the above equation, we obtain the 0th order conductivity as

$$\sigma_{xx}^{(0)} = e^2 \left( \frac{1}{2} v^2 \tau \right) \frac{\mu^2 - m^2}{2\pi v^2} \left( \frac{1}{\mu} \right) . \quad (9)$$

This expression can be compared with the conventional form $\sigma \sim e^2 D_0 N$, where $D_0 = v^2 \tau / 2$ is the diffusion constant and $N \sim \frac{\mu^2 - m^2}{2\pi v^2} \left( \frac{1}{\mu} \right)$ can be regarded as the density of carriers. The ladder diagram correction is given by

$$\sigma_{xx}^{(1)} = \frac{e^2}{2\pi v^2} v^2 \tau \frac{ (\mu^2 - m^2) \left( (A_0 + A_4)(\mu^2 - m^2) + 2A_2(\mu^2 + m^2) \right) \begin{pmatrix} 2A_0^2 (\mu^2 + m^2)(\mu^2 + 3m^2) + 2A_0 A_2 (\mu^4 - 2\mu^2 m^2 - 7m^4) \\ -A_0 A_4 (\mu^2 - m^2)(3\mu^2 + 5m^2) - 8A_2^2 m^2 (\mu^2 - m^2) \\ -2A_2 A_4 (\mu^2 - m^2)(\mu^2 - 3m^2) + A_4^2 (\mu^2 - m^2)^2 \end{pmatrix} }{ 4\mu \left( A_0 (\mu^2 + m^2) + A_2 (\mu^2 - m^2) \right) \left( A_0 (\mu^2 + 3m^2) - 4A_2 m^2 - A_4 (\mu^2 - m^2) \right)^2 }$$

(10)

The sum, $\sigma_{xx} = \sigma_{xx}^{(0)} + \sigma_{xx}^{(1)}$, gives the Drude conductivity up to the ladder diagram level. We find the conductance to be zero at the bottom of the conduction band, while at high doping level $\sigma_{xx}^{(0,1)}$ approach $e^2 (v^2 \tau / 2)(\mu / 2\pi v^2)$. Here we note that angular integrals are also energy dependent and except for $A_0$, all others go to zero at the high doping level, so that the net result is consistent to the earlier theoretical treatment [23] of pristine graphene.

Beyond the ladder diagrams level, the next-order modification to the conductivity is the MCD correction [5,9-12], shown in Figs. 1D and 1E. Such diagrams describe the so-called coherent backscattering effect [5,9-12]. In conventional 2DEG systems and in the presence of time-reversal symmetry, for each closed loop in propagating process, the clockwise propagating wave and the anti-clockwise propagating wave will interfere with each other. Such interference enhances the backward scattering, and therefore give divergent downward corrections to the diffusion constant [5,9-12] in 1D and 2D, thereby giving rise to the strong Anderson localization behavior [1,2]. In graphene systems, however, there is a Berry phase associated with the wavefunction, and if the Berry phase is $\pi$, then instead of constructive interference in the backward direction there will be destructive interference in the backward direction, thereby induces the so-called WAL [13-20] phenomenon. In the system described by Hamiltonian eq. (1), associated with a 0 to $\pi$ variation in the Berry phase [44,45], it is possible to have a WL to WAL transition. Of course due to the lack of a diffusion pole, the downward/upward correction will be non-divergent. Below we show the above scenario to be exactly the case. The MCD modification to conductivity can be written as:

$$\delta\sigma_{MCD} = \frac{e^2}{\omega}\int_{\mu-\omega}^{\mu}\frac{d\varepsilon}{2\pi}\sum_{\substack{k,k',i,j,l,m \\ \alpha,\beta,\gamma,\delta}} G^r_{k,i\alpha}J_{\alpha\beta}G^a_{k,\beta j}\Gamma_{k,k',i,j,l,m}G^r_{k',\gamma l}J_{\delta\gamma}G^a_{k',m\delta} \quad , \tag{11}$$

in which $\Gamma_{k,k',i,j,l,m}$ denotes the twisted MCD double-particle propagator. Under time-reversal invariance, MCDs can be transformed into ladder diagrams, which are easy to be written into the Bethe-Salpeter equation. The MCDs are shown diagrammatically in Figs. 1D, 1E, while the corresponding Bethe-Salpeter equation is given by

$$\Gamma_{k,k',i,j,l,m} = \sum_{k''} n_i^2 U_{k,k'',i,i}U_{k'',k,m,m}G^r_{k'',l,i}G^a_{k'',m,j}U_{k'',k',l,l}U_{k',k'',j,j} + \sum_{k'',r,s} n_i U_{k,k'',i,i}U_{k'',k,m,m}G^r_{k'',r,i}G^a_{k'',m,s}\Gamma_{k'',k',r,j,l,s} . \tag{12}$$

Details for the solution of Eq. (12) are given in Appendix C. Since the MCD correction $\delta\sigma_{MCD}$ cannot be calculated analytically, only the numerical result is shown. If the mean free path near the bottom of the conduction band is $l_p = \upsilon\tau \sim 30$ *nm*, the group velocity $\upsilon \sim c/300$, and the gap

(=2m) taken to be 53 *meV*, as the prediction of graphene/*h*-BN case, then the results are summarized in Fig. 2. In Fig. 2A, the total conductance, given by $\sigma_{total} = \sigma^{(0)} + \sigma^{(1)} + \delta\sigma_{MCD}$, is plotted as a function of the Fermi energy from the conduction band bottom. The variation of the Berry phase is also shown. The conductance curve has two regions, colored by red and blue, in which the blue region denotes negative $\delta\sigma_{MCD}$, while the red region denotes positive $\delta\sigma_{MCD}$. In Fig. 2B, the doping level dependence of $\delta\sigma_{MCD}$ is shown explicitly. As the MCD can be suppressed by the application of a magnetic field, thereby changing the resistance, this behavior of a WL to WAL transition directly implies a change of sign for the magneto-resistance, with zero magnetoresistance at the transition point.

We have also explored the concentration dependence of Coulomb impurities and examined its effect on the WL/WAL transition; the results are summarized in Fig. 3A. In the plots, the impurity concentration is characterized by the different values of the mean free path around the conduction band bottom, which is inversely proportional to the impurity density. We find the critical energy that distinguishes the WL and WAL to be independent on the mean free path. As mean free path increases, the critical energy of WL/WAL phase transition is at a constant value, ~46 meV. This can be attributed to the fact that the Berry phase is only dependent on energy. The corresponding critical Berry phase in this case ($m=26.5$ meV and $\upsilon = c/300$) is ~ $0.64\pi$. However, when the mean free path is very short, i.e., at the very high impurity concentrations, the MCD correction is seen to renormalize the total conductance to zero. Such a result indicates that not only WL is realized in such system, but also the Anderson localization. Since that $\sigma_{total} = \sigma^{(0)} + \sigma^{(1)} + \delta\sigma_{MCD}$, the realization of AL is because the MCD correction is negative and its absolute value is large enough to renormalize the total conductivity to zero. According to eq. (11), there are double integrations on k and k', respectively, and in calculation only the backward scattering is important, therefore we can replace the integral over k' with the integral over Q, where we set k= - k′, as shown in Fig. 1 (E). Physically, Q means the difference between the wave vectors of incoming waves and backscattered waves in a coherent backscattering path loop. Thus the integral over Q is a finite one, with the upper limit to be $1/l$, and the lower limit to be $1/L_{\text{dephasing}}$. Here $l$ means the mean free path and $L_{\text{dephasing}}$ is the dephasing length that waves lose their coherence. If integral over such whole Q region is large enough to renormalize the total conductivity to zero, then AL recovers. While in this case, actually we should increase our integral lower limit, to reduce the effect of MCD correction and keep the total conductivity to be zero but not negative. So the new lower limit of Q integral $1/\xi$ is higher than the original one $1/L_{\text{dephasing}}$, and it sets up a new length scale $\xi$, which can be considered as localization length in AL. With the parameter of graphene/h-BN, and with different mean free paths, we calculated corresponding possible AL localization lengths, and the results are shown in Fig. 3B. We can see the localization lengths increase with doping level, and finally diverge at the corresponding MIT critical points.

Finally we consider different masses and calculate all their transition behaviors, shown in Fig. 4. As shown in Fig. 4. A, we find that the WL/WAL transitions occur at almost the same critical Berry phase ~$0.64\pi$ and do not dependent on mean free path, which indicates the nature of such phase transition is due to the Berry phase variation. While for the AL/WL transition, since the MIT is dependent on the competition between the conductivity from ladder diagrams and MCD, in which the former depends on mean free path while the latter does not. So the AL/WL transition critical point depends on both mass term and mean free path. The MIT critical point versus mean free path plot is also illustrated in Fig. 4.B. We can see the left-down region that enclosed by the MIT critical line is the AL phase, and the up-right region denotes the

WL/WAL region, which can be considered as a metallic phase. An interesting feature can be observed, which is the production of intercepts on x-axis and on y-axis for different cases, i.e. the production of the MIT critical point as mean free path goes to zero and the maximum possible mean free path that AL can realize at the bottom of conduction band is approximately a constant, which is within the range ~ 8500±1000 meV·Angstrom.

In summary, we have theoretically investigated the electrical transport in 2D massive Dirac fermion systems, with long range Coulomb impurities. By calculating both the ladder diagram correction and the MCD correction to the Drude-conductance, we identified a continuous AL/WL/WAL transition as a function of the doping level away from the bottom of the conduction band, associated with the 0 to $\pi$ variation of the Berry phase. The uniform critical Berry phase for WL/WAL transition is about ~$0.64\pi$ despite different mass gaps and different mean free paths. In the other extreme of high impurity concentration, it is shown that the AL can be realized and mobility as well as localization length can be calculated. The existence of MIT with respect to doping level distinguishes the AL phase and the metallic phase in this 2D Dirac fermionic system, which is in contrast to the classical scaling theory, and also different from the 2D Weyl fermionic graphene systems. What's more, the MIT critical point is associated with the variation of Berry phase, which maybe a universal mechanism for possible occurrence of Localization/de-localization transition in material systems with topological non-trivial properties. Additional investigations on the scaling behaviors of this system are now ongoing and when accomplished, we'll be able to strictly calculate the scaling $\beta$ function to identify the localization to anti-localization MIT in real space. Combine the results from this manuscript and the results from scaling function calculations, the Berry phase induced MIT will be fully understood with different aspects.

**Acknowledgement**

The authors want to thank the financial support of Hong Kong RGC General Research Fund GRF16307114.

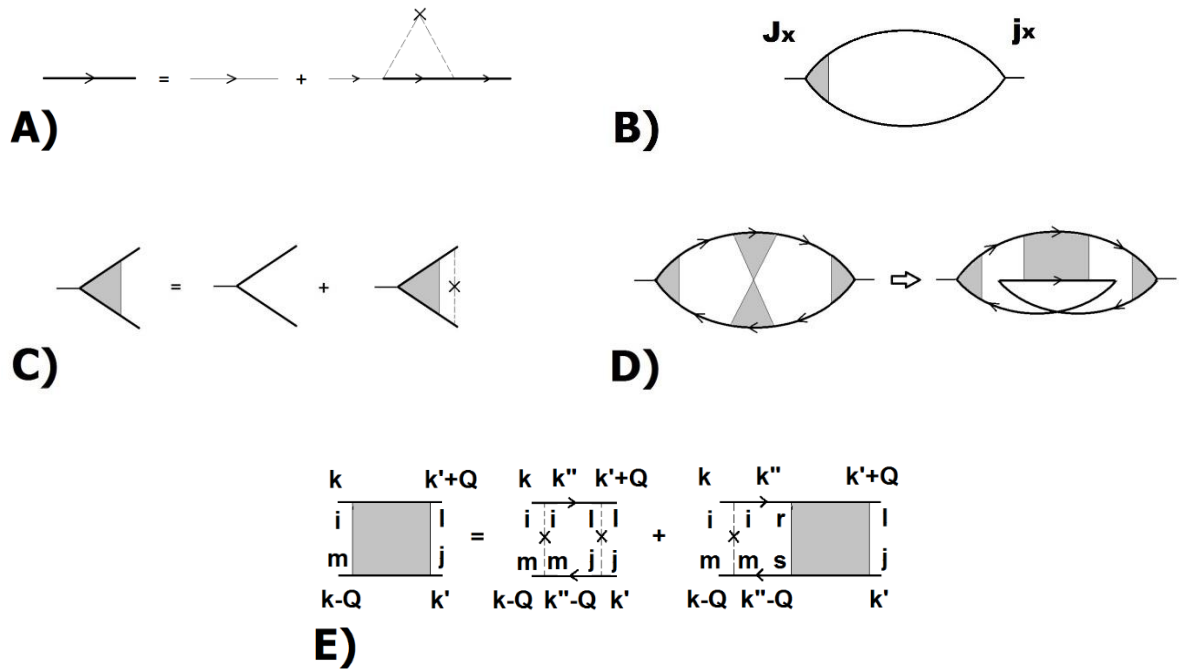

**Fig. 1. (A)** The correction to the single particle's Green function within the self-consistent Born approximation (SCBA). The diagram expresses the Dyson equation of the form $G_k = G_k^{(0)} + G_k^{(0)} \Sigma(k,\varepsilon) G_k$, in which the bold line denotes the corrected single particle Green function, while the thin line denotes the bare Green function. The dashed lines are the Coulomb interactions, and the cross denotes the impurity. **(B)** The diagram for the Drude conductivity, where the grey shading indicates the corrected current vertex shown in (C). **(C)** The ladder diagram correction to the current vertex, with the Bethe-Salpeter equation (Eq. (6)) expressed graphically. Here the grey shading indicates the self-consistent solution of the corrected current vertex $J_x(\vec{k},\vec{q},\omega)$ in Eq. (6). **(D)** The MCD correction to the Drude conductivity, and the transformation from the MCD to the twisted "Cooperons." This transformation changes the MCDs into the ladder diagrams that can be calculated with the Bethe-Salpeter equation. Here the grey shading indicates the corrected current vertices and the collection of all the MCDs, respectively. **(E)** The Bethe-Salpeter equation for the transformed MCD. Note that the lowest order diagram contains two impurity scattering vertices, in contrast to the ladder diagram where the lowest order diagram contains only one impurity scattering vertex.

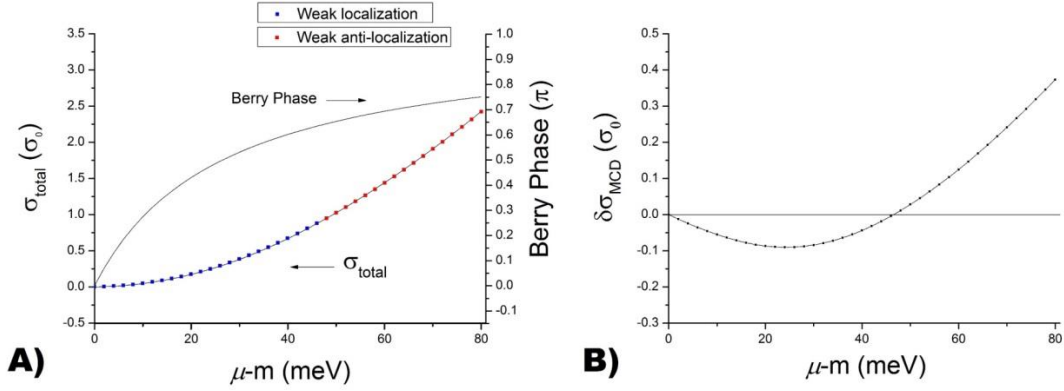

**Fig. 2.** **(A)** The calculated total conductivity per valley per spin and Berry phase versus energy in the conduction band, with mass $m=26.5$ meV and mean free path $l=30$ nm. In the plot the dotted line denotes the calculated conductivity $\sigma_{total} = \sigma^{(0)} + \sigma^{(1)} + \delta\sigma_{MCD}$ (the left vertical axis), and the solid line denotes the Berry phase (the right vertical axis). The blue section of the dotted line indicates the region of WL, and the red section indicates the region of WAL. The Berry phase increases from 0 at the bottom of the conduction band to $\pi$ when the doping level is far away from the band bottom. Here $\sigma_0 = e^2/\hbar$ denotes the quantum conductance. **(B)** The MCD correction to the conductance as a function of the Fermi level. The MCD correction is seen to change sign as the doping level increases. Negative correction denotes WL, while positive correction implies WAL. The transition point is about 46 meV from the bottom of the conduction band in this case, which corresponds to Berry phase $\sim 0.64\pi$.

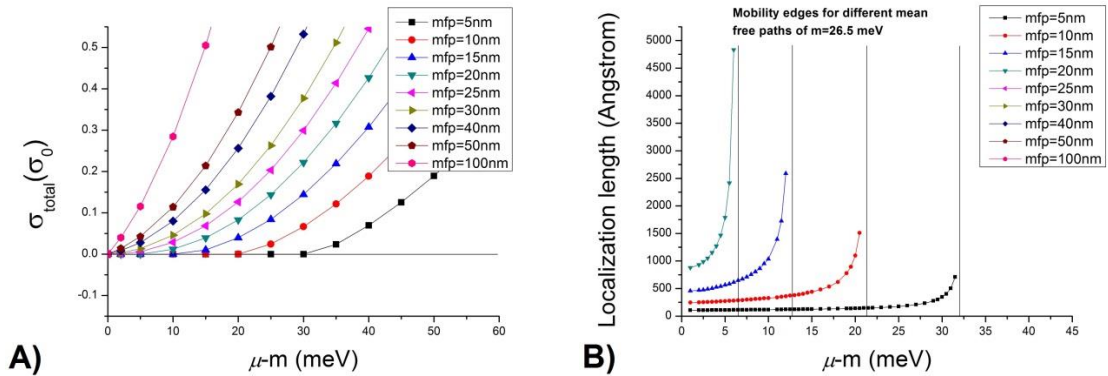

**Fig. 3.** **(A)** The total conductance of the system versus different mean free path in case $m=26.5$ meV. It is seen that for high impurity concentration (corresponds to a small mean free path), there can be a region near the conduction band bottom where the total conductivity is renormalized to zero, indicating presence of AL state. **(B)** The calculated localization lengths $\xi$ of different mean free paths as function of doping level for $m=26.5$ meV. When mean free path is larger than 20 nm AL cannot occur. The doping level that localization length diverges denotes the MIT critical point.

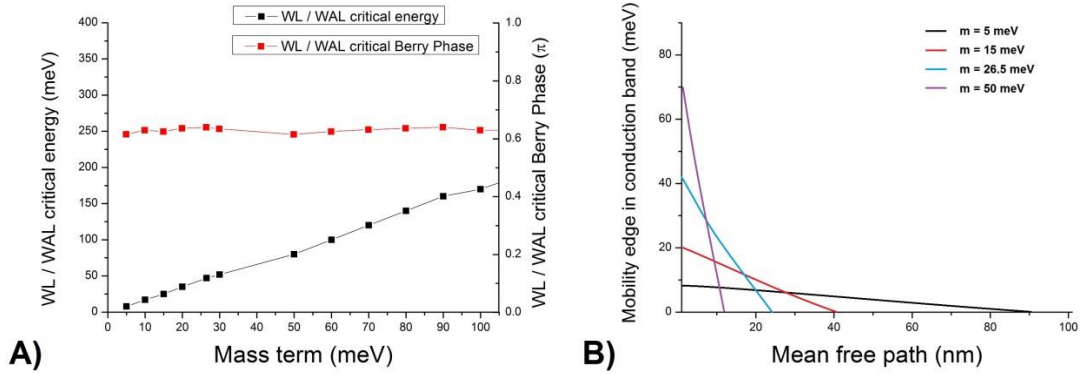

**Fig. 4. (A)** The WL to WAL critical energy versus different mass *m*, and corresponding critical Berry phase with respect to different masses. It's obvious the WL/WAL transition is governed by the variation of Berry phase, that the transition critical Berry phase is almost a constant ~ 0.64π. **(B)** the phase diagram for AL/WL transition. For different mass *m*, the MIT critical point, which distinguishes the insulating AL phase (left-down part) and the metallic WL/WAL phase (up-right part) varies with respect to mean free path.

## Appendix A: Self-energy due to Coulomb impurity scattering

Consider the self-energy shown in Eq. (4) in the main text:

$$\Sigma_{ij}(\vec{k},\varepsilon) = \int \frac{d\vec{k}'}{(2\pi)^2} \sum_{l,m=A,B} n_i u_{i,l}(\vec{k}-\vec{k}') G_{lm}(\vec{k}',\varepsilon) u_{m,j}(\vec{k}'-\vec{k}) , \qquad (A1)$$

and the corresponding Dyson equation for the single particle green's function:

$$G_k - G_k^{(0)} \Sigma(k,\varepsilon) G_k = G_k^{(0)} . \qquad (A2)$$

To calculate the self-energy, we substitute the expression for the Coulomb impurity potential $u_{kk'AA} = u_{kk'BB} = 2\pi e^2/\kappa p\varepsilon(p)$ and $u_{kk'AB} = u_{kk'BA} = 0$, as well as the $0^{th}$ order Green's function,

$$G^{(0)R/A}(\vec{k}',\varepsilon) \sim \frac{1}{2\varepsilon} \frac{1}{\varepsilon - \sqrt{v^2 k'^2 + m^2} \pm i\delta} \begin{pmatrix} \varepsilon + m & vk'e^{-i\theta'} \\ vk'e^{i\theta'} & \varepsilon - m \end{pmatrix},$$

into Eq. (A1). In the following, the relation $\int \frac{d\vec{k}'}{(2\pi)^2} = \int \frac{k'dk'}{2\pi} \int \frac{d\theta'}{2\pi}$ is employed to decouple the integral into an angular part and a radial part. To calculate the Coulomb potential integral, we notice that the denominator of $2\pi e^2/\kappa p\varepsilon(p)$ can be alternatively represented as $p\varepsilon(p) = p + ip_{TF} = 2k\sin\varphi + ip_{TF}$, where $\varphi = (\theta' - \theta)/2$. As a result, the self-energy can be expressed as:

$$\Sigma^{R/A} = \left(\frac{2\pi e^2}{\kappa}\right)^2 \int \frac{k'dk'd\theta'}{(2\pi)^2} n_i \frac{1}{\varepsilon - \sqrt{v^2 k'^2 + m^2} \pm i\delta} \begin{pmatrix} \frac{1}{4k'^2 \sin^2\varphi + p_{TF}^2} \frac{\varepsilon+m}{2\varepsilon} & \frac{1}{4k'^2 \sin^2\varphi + p_{TF}^2} \frac{vk'e^{-i\theta'}}{2\varepsilon} \\ \frac{1}{4k'^2 \sin^2\varphi + p_{TF}^2} \frac{vk'e^{i\theta'}}{2\varepsilon} & \frac{1}{4k'^2 \sin^2\varphi + p_{TF}^2} \frac{\varepsilon-m}{2\varepsilon} \end{pmatrix}. \qquad (A3)$$

From which it is straight forward to obtain

$$\Sigma^{R/A} = \mp i\pi n_i \left(\frac{2\pi e^2}{\kappa}\right)^2 \frac{\varepsilon}{2\pi v^2} \begin{pmatrix} \frac{\varepsilon+m}{2\varepsilon} A_0 & \frac{\sqrt{\varepsilon^2-m^2} e^{-i\theta}}{2\varepsilon} A_{-2} \\ \frac{\sqrt{\varepsilon^2-m^2} e^{i\theta}}{2\varepsilon} A_2 & \frac{\varepsilon-m}{2\varepsilon} A_0 \end{pmatrix} . \qquad (A4)$$

Here we have introduced the angular integrals $A_{2n} = \int \frac{2d\varphi}{2\pi} \frac{e^{i2n\varphi}}{(4k^2 \sin^2\varphi + p_{TF}^2)}$. The function to be integrated is $\frac{e^{i2n\varphi}}{(4k^2 \sin^2\varphi + p_{TF}^2)}$, and it takes the maximum magnitude at $\varphi=0$. In this case the function is completely determined by the Fermi momentum $p_{TF}$. Since $p_{TF} = 2\pi e^2 N_0/\kappa$, hence the self-energy due to long range Coulomb disorders is $\sim 1/N_0$, or $\tau \sim N_0$, the density of states. This feature is in consistent with the general conclusion for Coulomb impurities. The radial integral can be estimated by transforming the Green function into a delta function, which is the standard trick to obtain Eq. (A4).

Next we proceed to solve the corrected Green function from Eq. (A2). By substituting the result of self-energy, Eq. (S4) into Eq. (S2), we obtain the following equation set:

$$G^{R/A}_{1,1} = \frac{G^{R/A(0)}_{1,1} - \Sigma_{2,2}\left(G^{R/A(0)}_{1,1}G^{R/A(0)}_{2,2} - G^{R/A(0)}_{1,2}G^{R/A(0)}_{2,1}\right)}{1 - \left(G^{R/A(0)}_{1,1}\Sigma_{1,1} + G^{R/A(0)}_{1,2}\Sigma_{2,1} + G^{R/A(0)}_{2,1}\Sigma_{1,2} + G^{R/A(0)}_{2,2}\Sigma_{2,2}\right)}$$

$$G^{R/A}_{1,2} = \frac{G^{R/A(0)}_{1,2} - \Sigma_{1,2}\left(G^{R/A(0)}_{1,1}G^{R/A(0)}_{2,2} - G^{R/A(0)}_{1,2}G^{R/A(0)}_{2,1}\right)}{1 - \left(G^{R/A(0)}_{1,1}\Sigma_{1,1} + G^{R/A(0)}_{1,2}\Sigma_{2,1} + G^{R/A(0)}_{2,1}\Sigma_{1,2} + G^{R/A(0)}_{2,2}\Sigma_{2,2}\right)}$$

$$G^{R/A}_{2,1} = \frac{G^{R/A(0)}_{2,1} - \Sigma_{2,1}\left(G^{R/A(0)}_{1,1}G^{R/A(0)}_{2,2} - G^{R/A(0)}_{1,2}G^{R/A(0)}_{2,1}\right)}{1 - \left(G^{R/A(0)}_{1,1}\Sigma_{1,1} + G^{R/A(0)}_{1,2}\Sigma_{2,1} + G^{R/A(0)}_{2,1}\Sigma_{1,2} + G^{R/A(0)}_{2,2}\Sigma_{2,2}\right)}$$

$$G^{R/A}_{2,2} = \frac{G^{R/A(0)}_{2,2} - \Sigma_{1,1}\left(G^{R/A(0)}_{1,1}G^{R/A(0)}_{2,2} - G^{R/A(0)}_{1,2}G^{R/A(0)}_{2,1}\right)}{1 - \left(G^{R/A(0)}_{1,1}\Sigma_{1,1} + G^{R/A(0)}_{1,2}\Sigma_{2,1} + G^{R/A(0)}_{2,1}\Sigma_{1,2} + G^{R/A(0)}_{2,2}\Sigma_{2,2}\right)}$$

(A5)

The above can be simplified to the following form:

$$G^{R/A}(\vec{k},\varepsilon) = \frac{1}{2\varepsilon}\frac{\varepsilon + m\sigma_z + \upsilon\vec{\sigma}\cdot\vec{k}}{\varepsilon - \sqrt{\upsilon^2 k^2 + m^2} \pm i/2\tau_k}, \tag{A6}$$

with the scattering time given by

$$\frac{i}{2\tau_k} = -\frac{1}{2\varepsilon}\left((\varepsilon+m)\Sigma^R_{1,1} + (\upsilon k e^{-i\theta})\Sigma^R_{2,1} + (\upsilon k e^{i\theta})\Sigma^R_{1,2} + (\varepsilon-m)\Sigma^R_{2,2}\right) = i\left(\frac{2\pi e^2}{\kappa}\right)^2\frac{n_i}{4\upsilon^2}\left(\frac{\varepsilon^2+m^2}{\varepsilon}A_0 + \frac{\varepsilon^2-m^2}{2\varepsilon}(A_2+A_{-2})\right)$$

(A7)

## Appendix B: Vertex correction due to the ladder diagrams

The correction due to the ladder diagrams can be expressed in Eq. (7) in the main text, in which we have decoupled the Bethe-Salpeter equation onto different angular components:

$$J_x^{(0)} + J_x^{(1)}e^{i\theta} + J_x^{(-1)}e^{-i\theta} + J_x^{(2)}e^{i2\theta} + J_x^{(-2)}e^{-i2\theta}$$
$$= j_x + \int\frac{d\vec{k}'}{(2\pi)^2}n_i U_{k,k'}G^R(\vec{k}',\varepsilon)\left(J_x^{(0)} + J_x^{(1)}e^{i\theta'} + J_x^{(-1)}e^{-i\theta'} + J_x^{(2)}e^{i2\theta'} + J_x^{(-2)}e^{-i2\theta'}\right)G^A(\vec{k}'-\vec{q},\varepsilon-\omega)U_{k'-q,k-q}$$

(A8)

Here we define the integrals

$$\xi^n_{i,j,l,m} = \int \frac{d\vec{k}'}{(2\pi)^2} n_i U_{k,k',i,i} G^R_{i,l}(\vec{k}',\varepsilon) e^{in\theta'} G^A_{m,j}(\vec{k}'-\vec{q},\varepsilon-\omega) U_{k'-q,k-q,j,j} \quad , \tag{A9}$$

as the coefficients before the unknowns $J^n_x$ inside the Bethe-Salpeter equation. Substituting the expressions $u_{kk'AA} = u_{kk'BB} = 2\pi e^2/\kappa(2k\sin\varphi + ip_{TF})$ , $u_{kk'AB} = u_{kk'BA} = 0$ , and Eq. (A6):

$$G^{R/A}(\vec{k},\varepsilon) = \frac{1}{2\varepsilon} \frac{\varepsilon + m\sigma_z + \upsilon\vec{\sigma}\cdot\vec{k}}{\varepsilon - \sqrt{\upsilon^2 k^2 + m^2} \pm i/2\tau_k}$$ into Eq. (A9), we can calculate the integral analytically.

Again the integral can be divided into the angular and radial parts, as noted previously. In the angular part there are angular factors $e^{in\theta'}$ which arise from the decomposed corrected current vertex. And for the radial part, again the product of two Green functions can be written as a delta function, and therefore Eq. (A9) can be calculated. To be more specific, the common part in the product of two Green functions is:

$$\xi_{\text{kernel}} = \frac{1}{\varepsilon - \sqrt{\upsilon^2 k'^2 + m^2} + i/2\tau_{k'}} \frac{1}{\varepsilon - \omega - \sqrt{\upsilon^2(\vec{k}'-\vec{q})^2 + m^2} - i/2\tau_{k'}} = \frac{2\pi i}{\omega + i/\tau + \sqrt{\upsilon^2(\vec{k}'-\vec{q})^2 + m^2} - \sqrt{\upsilon^2 k'^2 + m^2}} \delta\left(\varepsilon - \sqrt{\upsilon^2 k'^2 + m^2}\right)$$

$$= 2\pi\tau \left(1 + i\omega\tau - \frac{i\upsilon^2 \tau k' q \cos\theta'}{\sqrt{\upsilon^2 k'^2 + m^2}} - \tau^2 \upsilon^2 \frac{\upsilon^2 k'^2 q^2 \cos^2\theta'}{\upsilon^2 k'^2 + m^2}\right) \delta\left(\varepsilon - \sqrt{\upsilon^2 k'^2 + m^2}\right)$$

(A10)

In the above expression we have translated the product of two Green functions' denominators into delta functions, and then Taylor-expand the expression with respect to $q$ and $\omega$.

Next we'd like to show an example of the integral, $\xi^0_{1,2,1,2} = \int \frac{d\vec{k}'}{(2\pi)^2} n_i U_{k,k',1,1} G^R_{2,1}(\vec{k}',\varepsilon) G^A_{1,2}(\vec{k}'-\vec{q},\varepsilon-\omega) U_{k'-q,k-q,2,2}$, which reads:

$$\xi^0_{1,2,1,2} = \int \frac{d\vec{k}'}{(2\pi)^2} n_i \left(\frac{2\pi e^2}{\kappa}\right)^2 \frac{1}{(4k^2 \sin^2\varphi + q_{TF}^2)}$$

$$\times 2\pi\tau \left(1 + i\omega\tau - \frac{i\upsilon^2 \tau k' q(e^{i2\varphi}e^{i\theta} + e^{-i2\varphi}e^{-i\theta})}{2\sqrt{\upsilon^2 k'^2 + m^2}} - \tau^2 \upsilon^2 \frac{\upsilon^2 k'^2 q^2(e^{i4\varphi}e^{i2\theta} + e^{-i4\varphi}e^{-i2\theta} + 2)}{4(\upsilon^2 k'^2 + m^2)}\right) \delta\left(\varepsilon - \sqrt{\upsilon^2 k'^2 + m^2}\right) \frac{\upsilon k' e^{i\theta'}}{2\varepsilon} \frac{\upsilon k' e^{-i\theta'}}{2\varepsilon}$$

(A11)

By taking the limit of $q,\omega \to 0$, we have:

$$\xi^0_{1,2,1,2} = 2\pi\tau n_i \left(\frac{2\pi e^2}{\kappa}\right)^2 \int \frac{k'dk'}{(2\pi)} \delta\left(\varepsilon - \sqrt{v^2 k'^2 + m^2}\right) \frac{v^2 k'^2}{4\varepsilon^2} \int \frac{d\theta'}{2\pi} \frac{1}{\left(4k^2 \sin^2\varphi + q_{TF}^2\right)} \tag{A12}$$

$$= 2\pi\tau n_i \left(\frac{2\pi e^2}{\kappa}\right)^2 \frac{1}{2\pi v^2} \frac{\varepsilon^2 - m^2}{4\varepsilon} A_0$$

We notice that the expression of scattering time has the form: $\frac{1}{\tau_k} = \left(\frac{2\pi e^2}{\kappa}\right)^2 \frac{n_i}{2v^2} \left(\frac{\varepsilon^2 + m^2}{\varepsilon} A_0 + \frac{\varepsilon^2 - m^2}{2\varepsilon}(A_2 + A_{-2})\right)$, therefore in this example calculation we have

$$\xi^0_{1,2,1,2} = \frac{\varepsilon^2 - m^2}{\left(2(\varepsilon^2 + m^2) A_0 + (\varepsilon^2 - m^2)(A_2 + A_{-2})\right)} A_0 \quad . \tag{A13}$$

With the scheme described above, all the integrals $\xi^n_{i,j,l,m}$ can be calculated. By substituting these integrals into the equation set Eq. (S8), we obtain the following solutions:

$$J_{x,1,1}^{(0)} = 0; \; J_{x,1,1}^{(2)} = 0; \; J_{x,1,1}^{(-2)} = 0; \; J_{x,1,1}^{(1)} = \frac{A_2(\varepsilon+m)\sqrt{\varepsilon^2 - m^2}}{4m^2(A_0 - A_2) + (\varepsilon^2 - m^2)(A_0 - A_2 + A_{-2} - A_4)} v;$$

$$J_{x,1,1}^{(-1)} = \frac{A_{-2}(\varepsilon+m)\sqrt{\varepsilon^2 - m^2}}{4m^2(A_0 - A_{-2}) + (\varepsilon^2 - m^2)(A_0 + A_2 - A_{-2} - A_{-4})} v; \; J_{x,1,2}^{(0)} = \frac{2A_0(\varepsilon^2 + m^2) + A_2(\varepsilon^2 - m^2) - A_{-2}(\varepsilon^2 + 3m^2) - A_{-4}(\varepsilon^2 - m^2)}{A_0(\varepsilon^2 + 3m^2) + A_2(\varepsilon^2 - m^2) - A_{-2}(\varepsilon^2 + 3m^2) - A_{-4}(\varepsilon^2 - m^2)} v;$$

$$J_{x,1,2}^{(1)} = 0; \; J_{x,1,2}^{(-1)} = 0; \; J_{x,1,2}^{(-2)} = 0 \quad J_{x,1,2}^{(2)} = \frac{A_4(\varepsilon^2 - m^2)}{A_0(\varepsilon^2 + 3m^2) - A_2(\varepsilon^2 + 3m^2) + A_{-2}(\varepsilon^2 - m^2) - A_4(\varepsilon^2 - m^2)} v;$$

$$J_{x,2,1}^{(0)} = \frac{2A_0(\varepsilon^2 + m^2) - A_2(\varepsilon^2 + 3m^2) + A_{-2}(\varepsilon^2 - m^2) - A_4(\varepsilon^2 - m^2)}{A_0(\varepsilon^2 + 3m^2) - A_2(\varepsilon^2 + 3m^2) + A_{-2}(\varepsilon^2 - m^2) - A_4(\varepsilon^2 - m^2)} v; \; J_{x,2,1}^{(1)} = 0; \; J_{x,2,1}^{(2)} = 0; \; J_{x,2,1}^{(-1)} = 0;$$

$$J_{x,2,1}^{(-2)} = \frac{A_{-4}(\varepsilon^2 - m^2)}{A_0(\varepsilon^2 + 3m^2) + A_2(\varepsilon^2 - m^2) - A_{-2}(\varepsilon^2 + 3m^2) - A_{-4}(\varepsilon^2 - m^2)} v; \; J_{x,2,2}^{(0)} = 0; \; J_{x,2,2}^{(2)} = 0; \; J_{x,2,2}^{(-2)} = 0;$$

$$J_{x,2,2}^{(1)} = \frac{A_2(\varepsilon-m)\sqrt{\varepsilon^2 - m^2}}{4m^2(A_0 - A_2) + (\varepsilon^2 - m^2)(A_0 - A_2 + A_{-2} - A_4)} v; \text{ and } J_{x,2,2}^{(-1)} = \frac{A_{-2}(\varepsilon-m)\sqrt{\varepsilon^2 - m^2}}{4m^2(A_0 - A_{-2}) + (\varepsilon^2 - m^2)(A_0 + A_2 - A_{-2} - A_{-4})} v.$$

(A14)

The above are the detailed calculations for the ladder diagram correction to the current vertex.

## Appendix C: Maximally crossed diagram (MCD) correction

The MCD correction to the conductivity can be expressed as Fig. 1D in the main text with the equation

$$\delta\sigma_{MCD} = \frac{e^2}{\omega}\int_{\mu-\omega}^{\mu}\frac{d\varepsilon}{2\pi}\sum_{\substack{k,k',i,j,l,m \\ \alpha,\beta,\gamma,\delta}} G^r_{k,i\alpha} J_{\alpha\beta} G^a_{k,\beta j}\Gamma_{k,k',i,j,l,m} G^r_{k',\gamma l} J_{\delta\gamma} G^a_{k',m\delta} \ , \tag{A15}$$

in which the corresponding Bethe-Salpeter equation for MCD is given by

$$\Gamma_{k,k',i,j,l,m} = \sum_{k''} n_i^2 U_{k,k'',i,i} U_{k'',k,m,m} G^r_{k'',l,i} G^a_{k'',m,j} U_{k'',k',l,l} U_{k',k'',j,j} + \sum_{k'',r,s} n_i U_{k,k'',i,i} U_{k'',k,m,m} G^r_{k'',r,i} G^a_{k'',m,s} \Gamma_{k'',k',r,j,l,s} \ . \tag{A16}$$

Here k'~ -k+q as defined in ref. [5]. Equation (A16) is hard to solve, but we have an alternative equation:

$$e^{in\theta}\Gamma_{k,k',i,j,l,m}e^{in'\theta'} = e^{in\theta}\sum_{k''} n_i^2 U_{k,k'',i,i} U_{k'',k,m,m} G^r_{k'',l,i} G^a_{k'',m,j} U_{k'',k',l,l} U_{k',k'',j,j} e^{in'\theta'} + e^{in\theta}\sum_{k'',r,s} n_i U_{k,k'',i,i} U_{k'',k,m,m} G^r_{k'',r,i} G^a_{k'',m,s} \Gamma_{k'',k',r,j,l,s} e^{in'\theta'}$$

(A17)

in which the left hand side and right hand side both contain the same unknowns, $e^{in\theta}\Gamma_{k,k',i,j,l,m}e^{in'\theta'}$. So Eq. (A17) is actually an equation set, by solving which we can obtain the MCD correction to the conductivity. The 1$^{st}$ term on the right hand side, which contains 4 Coulomb interactions, generate terms inside the angular integrals like

$$U_{k,k'',i,i} U_{k'',k,m,m} U_{k'',k',l,l} U_{k',k'',j,j} \sim \frac{1}{\left(4k^2\sin^2\varphi + p_{TF}^2\right)\left(4k^2\sin^2\varphi' + p_{TF}^2\right)} \ , \tag{A18}$$

if subscripts $i,j,l,m$ take the proper values. The angular integral is therefore complicated. However, Eq. (S18) can be decomposed into the sum of two terms:

$$\frac{1}{\left(4k^2\sin^2\varphi + p_{TF}^2\right)\left(4k^2\sin^2\varphi' + p_{TF}^2\right)} = \frac{1}{4k^2 + 2q_{TF}^2}\left(\frac{1}{4k^2\sin^2\varphi + p_{TF}^2} + \frac{1}{4k^2\sin^2\varphi' + p_{TF}^2}\right), \tag{A19}$$

in which each term can be integrated (in MCD the fact that $\theta' \sim \theta + \pi$ is used; and since $\varphi = (\theta'' - \theta)/2$ and $\varphi' = (\theta' - \theta'')/2$, hence we have $\sin^2\varphi + \sin^2\varphi' \sim 1$). With this trick we can continue the calculation of angular integrations. The full solution of Eq. (A17) can only be achieved numerically, and the results are given in the main text.